# The ImmersaDesk3 - Experiences With A Flat Panel Display for Virtual Reality

Dave Pape, Josephine Anstey, Mike Bogucki, Greg Dawe, Tom DeFanti, Andy Johnson, Dan Sandin
Electronic Visualization Laboratory (EVL)
University of Illinois at Chicago
851 S. Morgan St., Room 1120
Chicago IL 60607-7053
{ pape, anstey, mbogucki, dawe, tom, aej, dan } @evl.uic.edu

**Abstract**

In this paper we discuss the design and implementation of a plasma display panel for a wide field of view desktop virtual reality environment. Present commercial plasma displays are not designed with virtual reality in mind, leading to several problems in generating stereo imagery and obtaining good tracking information. Although we developed solutions for a number of these problems, the limitations of the system preclude its current use in practical applications, and point to issues that must be resolved for flat panel displays to be useful for VR.

**1. Introduction**

As a continuation of our development of VR display devices that include the CAVE™ [1] and the ImmersaDesk™ [2], we are interested in designing a range of VR devices for a single user, including a desktop VR system. The ImmersaDesk is sized for a laboratory, but is still too large for a typical faculty office or cubicle. Its size was determined partly because we wanted to present a wide angle of view, but also because it is based on a CRT projector, and available projection technology limits how small the screen can be (approximately 6' diagonal).

Contrariwise, a number of commercial video display manufacturers are beginning to develop flat panel video display systems on a large enough scale to be used in a desktop VR system – up to 42" diagonally. As an initial test of the feasibility of flat panel screens for VR, we created the ImmersaDesk3 prototype using a commercial plasma display panel mounted on a standard office desk (Figure 1). The goal of this system was to create a head-tracked, stereoscopic desktop display in a space conserving, portable package, one that would be useful in an ordinary office. The basic system is a fishtank VR display [3], but in comparison to typical workstation monitors, the latest flat panel screens provide a much larger display while requiring less desktop space.

Our interest in flat panel display goes well beyond their use for small VR systems. These displays have the potential to be a major boon to projection-based virtual reality in general. Current systems, such as the CAVE, have a very large footprint; a 10 by 10 by 10 foot CAVE can actually require a 30 by 20 by 14 foot space when rear projection distances are included, even when these are folded with mirrors. If a CAVE could be built of flat panel displays, the savings in architectural requirements would be significant. Smaller scale projection-based systems, such as the Responsive Workbench [6] and ImmersaDesk, would also benefit from more space-efficient displays.

However, while putting the ImmersaDesk3 together, we encountered a number of limitations with the current flat panel displays. This paper gives an overview of the design of our system, and details the various problems we have found; these problems need to be addressed if this technology is to be truly useful in VR.

**2. Design of the ImmersaDesk3**

**2.1 Display Hardware**

Several different technologies for flat panel displays are being researched by manufacturers. These technologies include liquid crystal, ferro-electric liquid crystal, plasma panel, and light emitting diode displays. In our investigations of flat panels ([4]) we determined that, although liquid crystal (LCD) panels have high resolution

(1280x1024 or better), they are too slow to be used for stereo. Ferro-electric liquid crystal displays have the benefits of LCDs, but should be fast enough for field-sequential stereo imaging; however, they are presently only available in developer kits. Light Emitting Diode displays are bright and potentially borderless, but have very low resolution (e.g. 208x272 and 320x192).

Plasma panels were the only commercially available displays which met our needs for the ImmersaDesk3 – they were available in large desktop sizes (up to 42" diagonally), and at moderate cost (approximately $10,000). We considered that the plasma panel had a high enough resolution and was probably fast enough to do stereo. We chose to build our system around the Fujitsu PDS4201U-H Plasmavision display panel. The Plasmavision has an active display area of 36 x 20 inches (in a 16:9 aspect ratio); the entire panel is 41 x 25 x 6 inches. The panel weighs 80 lb.

The workstation driving the display was a deskside Silicon Graphics Onyx Infinite Reality.

**2.2 The Desk**

We mounted the Plasmavision on a modified office desk. To accommodate different applications, and for greater flexibility, we wanted to be able to position the screen vertically (perpendicular to the desktop), horizontally (flat on the desktop), or at an angle in between. The panel is too heavy for users to shift easily, so we mounted it on hydraulic supports with a hand crank to adjust the angle (Figure 2).

**2.3 Stereo Video**

Stereoscopic display in projection-based systems such as the ImmersaDesk is based on interleaving images for the left and right eyes in a single video signal. The video signal is typically very high bandwidth, being workstation resolution (1280x1024) at 96 to 120 Hz, which yields 48 to 60 Hz per eye. Liquid crystal shutter glasses are synchronized with this video signal to mediate the images. The shutter glasses are normally wireless, using infrared emitters to trigger them. We encountered a number of problems as we attempted to display stereo on the plasma panel.

Although the Plasmavision will accept a wide range of video frequencies as input, the panel re-scans the input video signal, and always displays a 30 Hz interlaced (i.e. NTSC quality) image. If our rendering software is to generate useable stereoscopic images, the signal that comes from the Onyx workstation has to be the same as that displayed on the Plasmavision screen, so we were restricted to an NTSC format display, with left and right eye images rendered in alternate video fields (halves of the video frame).

Since the NTSC video format was the only signal compatible with the plasma-panel, we felt that the 3DTV Model 2001 3D-Theatre shutter glasses might work with this display [7]. The 3DTV system is designed to be used in conjunction with a television / VCR setup, which uses a NTSC signal. This entire system is driven by a videocassette that provides a prerecorded stereo signal and 3D movie. The stereo signal is acquired from the video by an interconnection between the VCR's video-out and the television's video-in.

In order to drive the 3DTV unit with the SGI we needed to acquire some sort of sync signal. Tapping the 'sync-on-green' from the SGI provided us with this signal. Despite all of our attempts in trying to use the 3DTV shutter glasses, we could only achieve partial success with the unit. We observed that the stereo pairs tended to 'drift' from being separate to a fully fused stereo image. This led us to conclude that the signal going into the display was constantly being modified and displayed at a different rate.

It was also possible that the LCD shutters in the 3DTV unit were not refreshing fast enough. An alternative path was to try a pair of CrystalEyes shutter glasses made by StereoGraphics. These glasses can refresh at a rate of up to 120 Hz, whilst the latter unit can only go as high as 60Hz. Both types of glasses acquire their sync signal similarly via a small emitter box. However, the Crystal Eyes glasses are wireless; all the driver electronics are on the glasses. The sync signal is sent to them via an infrared beam. We discovered that the plasma-panel emits infrared radiation within the same spectrum as that used by the glasses causing them to 'misfire'. We tried using different infrared filters to block out some of the emitted radiation but met with no success. To use them at all we would have to modify the glasses and send them the sync signal on a wire.

But first we had to obtain an accurate sync signal. It was proposed that we might be able to extract the signal from somewhere within the display electronics. However, we were not able to obtain any pertinent information either from the documentation or from the Fujitsu representatives. A quick check within the unit was no more illuminating. Since schematics were not available, we were unable to determine the proper point on the circuit board where we might find a usable signal. Too much experimentation risked damaging the unit.

Finally, we decided to extract the sync signal via an optical pickup on the front of the display, and use a wired method to drive a pair of Crystal Eyes. A high-speed photodiode was place in the lower left-hand corner of the plasma-display (see Figure 3); a small white patch was drawn in the odd image fields while a black patch was drawn in the even image fields. This produced a black and white flickering which signaled the alternation of video fields.

The signal that came off of the photodiode was analog, therefore we needed to convert it to digital for the shutter glasses. The analog signal was fed into a Schmitt trigger; for every analog pulse that rose above a 4.76-volt threshold, the Schmitt trigger would fire a digital pulse. Anything that was sustained beyond the 4.76 threshold would be ignored. The digital pulse that came out of the trigger was then inverted, by a hex inverter.

At this point we had a digital signal that appeared to be periodic (Figure 4). To clean and modify this signal three monostable multivibrators were linked in series. The first multivibrator allowed us to create a digital pulse with a 16 millisecond delay that would overlook the rest of the pulses after the first initial pulse. The other two multivibrators allowed us to adjust the wavelength and duty cycle. This gave us a usable signal that was fed into a driver for the shutter glasses. Since the shutter-glasses driver needed to amplify the digital signal to roughly 40 volts, we needed to add a buffer before this stage to help control the current being drawn from the previous stages (Figure 5).

We modified a pair of CrystalEyes glasses by bypassing the internal electronics and feeding wires directly to the LCD's. The entire electronics package was easily housed in a box the size of a small book. In order for this device to work, the circuit needed to be calibrated by hand; we accomplished this by using an oscilloscope to observe the signals, while we also watched the plasma-display for stereo fusion.

### 2.4 Tracking

The ImmersaDesk3 used an Ascension PC Bird electromagnetic tracking system. The transmitter was mounted on the underside of the desktop near the front (just above a seated user's legs). We first tried using the lower cost Ascension Spacepad system, but the Plasmavision generated so much electromagnetic noise that it induced large positional errors. The PC Bird is not as susceptible, although if a sensor is very close to the screen, the tracking becomes noisy; this can make direct interaction with virtual objects awkward. Also, if the Plasmavision screen is between the transmitter and the receivers, the tracking fails. The location we chose for the transmitter kept it out of the user's way without being occluded by the panel.

Nevertheless, the tracking was still problematic. Further investigation has suggested that the cooling fans in the plasma panel are the most likely culprits for tracker inaccuracy.

### 2.5 Input Devices

We used EVL's standard wand as our primary input device; the wand has three buttons and a small pressure sensitive joystick, and has a tracking sensor attached. In some cases, we also used the workstation mouse as a second input device, to experiment with mixing 3D tracked and desktop interaction styles. The mouse was treated as a tracked device; the 3D position reported for it was the real world position of the pointer on the screen, computed from the pointer's graphical screen position and the measured geometry of the plasma panel. With this system, we could include menu and widget controls as 3D objects in the virtual world, but interact with them with the mouse, like a normal 2D desktop application, while also interacting with the rest of the application with the 3D trackers.

## 3. Testing the Desk

The ImmersaDesk3 was demonstrated at the SIGGRAPH '98 Emerging Technologies venue, in the installation "Guerilla VR". We showed seven VR applications, three of which were networked between the ImmersaDesk3 and an ImmersaDesk2 also on the show floor.

Part of our intention was to show the increasing flexibility, portability and accessibility of VR devices. Every day this desk was rolled to three different locations in the Orlando Convention Center. Within minutes the device was set up and demonstrating networked VR. This meant mounting the desk on wheels. The plasma panel remained mounted to the desk at all times, along with the PC to control the tracking system. The deskside Onyx rolled separately.

Since standard CAVE-based applications can run on the ImmersaDesk3 we were able to evaluate a variety of applications for their suitability on the desk. When the user was sitting comfortably at the desk, the Plasmavision did not fill his field of view, but was large enough to feel fairly immersive. Virtual spaces designed to surround the user in the CAVE (such as architectural spaces) did not work as well as spaces or objects designed to be viewed from the outside in. The ImmersaDesk3 was particularly effective in the networked demonstrations. In one educational application, a teacher in front of the plasma panel taught 4D math to a student on the Immersa-Desk2. Collaborative tasks were shared between the two desks. Using a menu and a mouse, a participant at the ImmersaDesk3 directed the experience of the user on the ImmersaDesk2.

The application developers and many users were very pleased and with the brightness and color quality of the plasma panel, in comparison to the CRT video projectors of the ImmersaDesk and CAVE. Its size gave a much larger angle of view than a conventional monitor, yet it fit easily on a desktop, something a 42" CRT monitor or a projection system cannot do. The Plasmavision was rugged. It survived the trip from Chicago to Orlando and back mounted on the desk, and survived repeatedly rolling it through crowds at SIGGRAPH.

## 4. Problems Encountered

Several of the problems we encountered with the Plasmavision were described above; we summarize them here, along with other significant problems.

The Plasmavision is electromagnetically noisy, causing it to interfere with typical VR tracking systems. It also emits a lot of infrared light, which interferes with standard wireless stereo shutter glasses.

The Plasmavision will only display a 30 Hz, interlaced, NTSC resolution image. The NTSC television standard uses a 30 Hz interlaced signal because it gives an effective 60 Hz (60 fields per second) refresh rate, which is fast enough that viewers do not see the image flicker. But when this is used for a stereoscopic display, each eye is only seeing a 30 Hz signal, and the flicker is very noticeable; prolonged exposure can give many users headaches.

Using the NTSC field-interleaved format for stereo yields only 640 x 240 pixel resolution for each eye's image. This is much lower than typical workstation resolution, and is insufficient for highly detailed applications. However, the graphics for some applications do run much faster at this reduced resolution.

While using the display with the shutter glasses, we found that the red and green phosphors do not decay quickly enough. When we looked at a stereo test pattern, which displays separate red, green, and blue color bars for each eye, only the blue bar was sufficiently extinguished; the red and green bars were still visible to the 'wrong' eye. The left eye can see the right eye's red and green bars dimly, and vice versa. Examining the signal with the photocell and an oscilloscope, with different colors displayed, verified that the apparent ghosting was due to long decay times in red and green. This results in significant cross-talk in most applications, to the degree that it is difficult for many users to properly fuse the stereo images. In an informal test of 16 users at SIGGRAPH, we found that 25% of them could fuse the full-color images, while 50% could only fuse the images when the red and green channels were disabled, so that the images were just shades of blue; 25% of the users could not fuse the images at all. Interestingly, for all of the users who had to view blue-only images to fuse, once they had successfully fused the images, we could turn the red and green back on, and they were able to 'hold' the stereo.

## 5. Conclusions

Current plasma panel technology has severe limitations as a stereo display device for projection-based virtual reality systems. The inability to easily sync the plasma panel to shutter glasses and the red/green phosphor decay problem preclude clear stereo. The low resolution and 30Hz frame rate also prevent current panels from being serious contenders in this field. Although flat panels could significantly save space, larger display systems would need larger panels, or would have to tile several of them. Current plasma panels have borders that prevent seamless tiling.

Nevertheless, both the concepts of a wide field of view, desktop VR system and space-saving flat panel technology for large displays such as the CAVE are still very appealing. We will therefore continue to investigate the most appropriate flat panel technology as it evolves.

**Acknowledgments**

The virtual reality research, collaborations, and outreach programs at EVL are made possible through major funding from the National Science Foundation, the Defense Advanced Research Projects Agency, and the US Department of Energy; specifically NSF awards CDA-9303433, CDA-9512272, NCR-9712283, CDA-9720351, and the NSF ASC Partnerships for Advanced Computational Infrastructure program.

CAVE and ImmersaDesk are trademarks of the Board of Trustees of the University of Illinois. Plasmavision is a trademark of Fujitsu General Limited.

**References**

[1] C. Cruz-Neira, D. J. Sandin, T. A. DeFanti. Surround-Screen Projection-Based Virtual Reality: The Design and Implementation of the CAVE. Computer Graphics (SIGGRAPH '93 Proceedings), Vol. 27, pp. 135-142, August 1993.

[2] M. Czernuszenko, D. Pape, D. Sandin, T. DeFanti, G. L. Dawe, and M.D. Brown. The ImmersaDesk and Infinity Wall projection-based virtual reality displays. Computer Graphics, Volume 31, Number 2, pp. 46-49, May 1997.

[3] M. Deering. High resolution virtual reality. Computer Graphics (SIGGRAPH '92 Proceedings), 26(2), pp. 195-202, July 1992.

[4] T. DeFanti, D. Sandin, G. Dawe, M. Brown, M. Rawlings, G. Lindahl, A. Johnson, J. Leigh. Personal Tele-Immersion Devices. In Proceedings of 7th IEEE International Symposium on High Performance Distributed Computing, pp. 198-205, July 1998.

[5] Fujitsu General Limited. Fujitsu 42" Wide Plasmavision User's Manual.

[6] W. Krueger, B. Froehlich. Visualization Blackboard: The Responsive Workbench (virtual work environment). IEEE Computer Graphics and Applications, 14(3), pp. 12-15, May 1994.

[7] 3DTV Corporation. http://www.stereospace.com/.

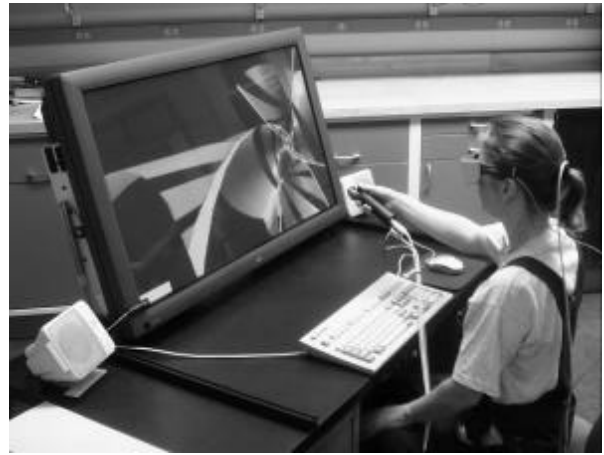

Figure 1. The ImmersaDesk3

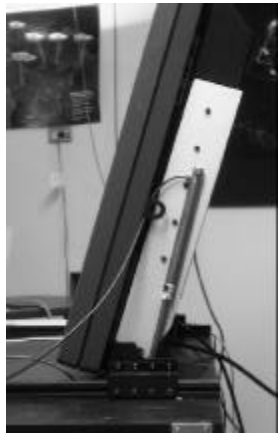

Figure 2. Hydraulic supports

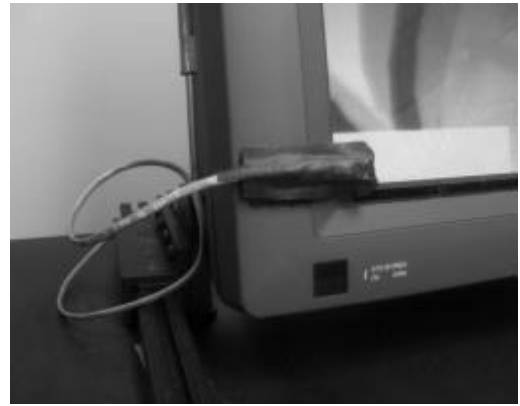

Figure 3. Photocell for stereo sync

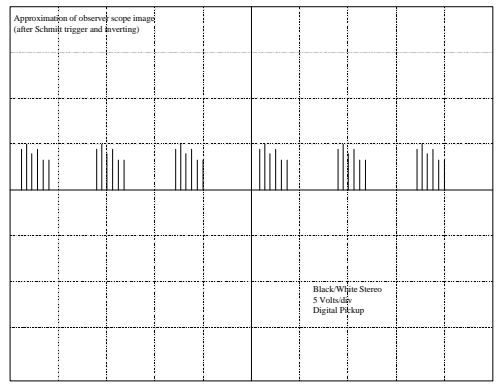

Figure 4. Stereo phase signal

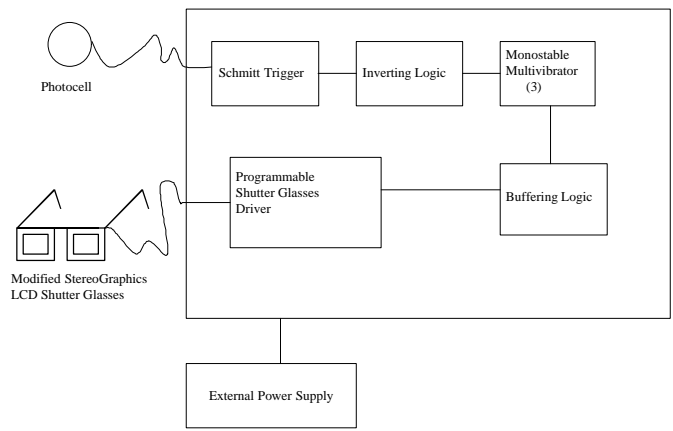
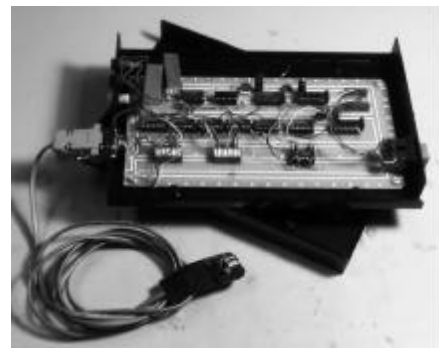

Figure 5. Stereo sync extraction box